\documentstyle[12pt]{article}
\begin{document}
\title{ Coarse grainings and irreversibility in quantum field theory} 
\author {C. Anastopoulos \\ Theoretical Physics Group, The Blackett Lab. \\
Imperial College \\ E-mail : can@tp.ph.ic.ac.uk \\}
\date { October 1996}
\maketitle
\begin{abstract}
In this paper we are interested in the studying coarse-graining in field
theories using the language of quantum open systems. Motivated by the
ideas of  Calzetta and Hu \cite{CaHu} on correlation histories we
employ the Zwanzig projection technique to obtain evolution equations
for relevant observables in self-interacting scalar field
theories. Our coarse-graining operation consists in concentrating
solely on the evolution of the correlation functions of degree less
than $n$, a treatment which corresponds to the familiar from
statistical mechanics truncation of the BBKGY hierarchy at the n-th
level. We derive the equations governing the evolution of mean
field and two-point functions thus identifying the terms corresponding to
dissipation and noise. We discuss possible applications of our
formalism, the emergence of classical behaviour and the connection to
the decoherent histories framework. 
\end{abstract}

 \renewcommand {\theequation}{\thesection.\arabic{equation}}
\let \ssection = \section
\renewcommand{\section}{\setcounter{equation}{0} \ssection}
\pagebreak

\section{Introduction}
\paragraph{Motivation}
Quantum field theory has a rich structure, which manifests itself in
the possibility of describing the same field system from diverse
points of view. Hence, depending on the problem of interest one could
focus for instance on the Hamiltonian, the statistical or the particle
aspects of the quantum field. This potentiality for description  within
different frameworks, inherent in quantum 
 field
theory is the cause of its large domain of applications, 
but is also a source of interesting  questions.  
\par
 The more important one is to identify the level of observation in a
field theory or, putting it another way, 
  what  an actual observer measures in a
quantum field. The answer to this question is not easy and it is clear
that the level of observation cannot be fixed uniquely. Unlike
non-relativistic quantum mechanics where one is essentially measuring
phase space quantities for a particle system spatially localized, in
quantum field theory local measurements contain only a very small
portion of information about the state of the field. A local observer,
will for instance be able to record only the mean field and the higher
order correlation functions are inaccessible to him. Therefore, for
most 
of the possible configurations, he might lose all sense of
predictability for the field observables. 
\par
This is closely connected with the problem of the classical limit of
field theories. In the context of the decoherent histories approach to
quantum mechanics, the classical domain corresponds to a set of
coarse-grained and non-interfering histories, from which one can
obtain almost deterministic equations for a class of observables
\cite{GeHa}.  In
our case there is a large number of such classes. 
 At low energies
one might consider the particle-like behaviour of the fields and obtain
in the classical limit a theory of interacting non-relativistic
particles. Or one could concentrate on phase space histories, to see the
extent to which QFT behaves as a Hamiltonian system. Or even consider 
histories of quantities like energy and momentum density and obtain
a classical hydrodynamics description. 
\par
The above issues are also of value for  early Universe
cosmology. The transition from quantum to classical is of great
importance in models of inflation since many of its predictions are
based on the fact that the long wavelength modes of the inflaton
exhibit classical behaviour. When considering the non-equilibrium
dynamics of fields (mainly for the study of phase transitions) the
first point needed to be settled is what are the variables we should
concentrate, that contain the relevant information for the problem in
hand.
\par
  The notion of natural coarse-graining in field theories is also
important in the
context of field theories in curved spacetime. For it is only one
quantity that actually governs the backreaction dynamics of spacetime:
 the expectation value of the field energy-momentum tensor
(essentially constructed from the two-point correlation functions in
the case of free fields).
\par
To address these problems a number of techniques from non-equilibrium
statistical mechanics has been employed with varying degree of
success: the Feynman-Vernon influence functional technique
\cite{FeVe,CaLe,HuMa} and  the close time path formalism \cite{CaHu3, CaHu2}. It is the
aim of this paper to exhibit the use of another powerful technique of
statistical mechanics in a field theoretic context: 
the Zwanzig projection method (for a review see \cite{RaMi, Zeh,
LiWe}). 
The great advantage of this method lies
in its wide range of possible applications: for any choice of coarse-graining
 it can be applied once we are able to identify the coarse graining
operation with an indempotent map on the space of states. Our choice
of coarse-graining is motivated by the ideas of Calzetta and Hu
\cite{CaHu} on the truncation of the Schwinger-Dyson hierarchy of
n-point functions.
\par
But before discussing the approach we adopt at this paper, we find
meaningful to give a short discussion on possible  choices for
coarse graining.   
\paragraph{Coarse grainings}  There are two important constraints one might
impose on our possible choices for coarse-graining: 
{\it naturality} and {\it Lorentz covariance}.
To see what we mean by naturality, let us consider cases of typical
coarse grainings in standard non-equilibrium statistical
mechanics. A typical situation is to separate relevant and
irrelevant observables according to the order of magnitude of some
physical parameter characterizing them. Hence we can for instance
average out the effect of ``fast'' variables (evolving within very
short timescales) or trace out the contribution of particles with
the smaller masses (as is the case in quantum Brownian motion). Such a
separation of scales, while quite common  in non-relativistic many
particle systems is rare in relativistic quantum field theory. If
possible it would involve a fine tuning of the the coupling constants
and masses of the field systems as well as the imposition of a particular
initial condition. In a generic systems it is unlikely that such
``autocratic '' coarse grainings can emerge naturally \cite{CaHu}.
\par
The requirement of Lorentz covariance, though it can be relaxed in a
number of situations (like for instance when non-relativistic matter
is present \cite{AnZu}) is of great importance both for the
cosmological applications and the emergence of classical behaviour for
the field variables. For when we try to study a field system for first
principles, there is no natural way a non-Lorentz invariant quantity
can be introduced in our schemes. Hence, for instance,  a coarse
graining taking the 
form of a high momentum cut-off for the field modes should not be
considered as fundamental but rather as emerging from the full
dynamics of the theory under particular circumstances.
\par
Besides those two {\it a priori} criteria for our choice of
coarse graining, there is an equally important one that can be considered
only {\it a posteriori}, that is after we have identified the dynamics
of the relevant variables. This is the requirement of {\it persistent
predictability} for the evolution equations. In the language of the
decoherent histories approach it states that histories of relevant
observables ought to form  a quasiclassical domain. This means that
the evolution equations have to be approximately dynamically
autonomous \cite{Zeh} (even though we cannot expect to obtain
Markovian behaviour). This again implies that the noise due to the
irrelevant part of the field, though sufficient to decohere the
histories of relevant observables is  weak enough to allow a degree of
predictability \cite{GeHa}. In general, this is expected to be
possible only for a small class of initial states of our system (or
the Universe for cosmological applications). This is a fact that
we will verify in our analysis.
\paragraph{Truncation of  the Schwinger-Dyson hierarchy}
The coarse-graining operation we shall examine, is one proposed by
Calzeta and Hu \cite{CaHu}. They remarked on the similarity of the
chain of Dyson equations linking each Green function to others of
higher order with the BBKGY hierarchy of correlation functions in
classical statistical mechanics. Since the set of expectation values
of field products contains all information about the state of the
field, a truncation in the chain of Green function will form a natural
coarse-graining operation and the lower order $n$-point functions
will be our relevant observables. The authors then proceed to compute
the effective equations of motion from a master effective action 
using a generalizaion of the close
time-path formalism. The important feature of these equations is the
presence of correlation noise, which under particular conditions may
guarantee decoherence of the ``correlation histories''.
\par
The  truncation of the Schwinger-Dyson
hierarchy does satisfy the conditions of Lorentz covariance and
naturality for the choice of the coarse graining operation. First,
this choice of coarse graining is closer to actual measurements of the
quantum field since any finite measurement device cannot obtain
information about arbitrarily high orders of correlation. Actually,
a local observer might be expected to monitor only the mean field
values. Second, being an intrinsically justifiable division between
relevant and irrelevant observables it can be applied to a
wide variety of systems, without the need to recourse to special 
arguments for each particular case. Third, it seems promising when trying to
consider evolution of hydrodynamic quantities since quantities like
energy and momentum density can be obtained through the knowledge of
low order correlation functions. In particular, when dealing with the
backreaction problem in curved spacetime, truncation of the hierarchy
at the level $n = 2$ might give interesting results since the
energy-momentum tensor determining backreaction can be determined
through the knowledge of 2-point functions.
\par
As far as the third requirement of  predictability is
concerned, we need to have a detailed calculation of the dynamical
evolution of the relevant observables. Still, it is important to note,
that the classical behaviour of the two point correlations observed at
later stages, gives us at least a
hint for the possibility of an initial condition such that the
dynamics of observables obtained from a truncation at the level $n =2$
are approximately autonomous.
\paragraph{The Zwanzig method}
To obtain the evolution equations for the relevant observables, we are
going to utilize, as mentioned earlier,  the Zwanzig projection technique. 
There is a number of reasons for believing that this provides an
important calculational tool when dealing with the above isssues:
\\
1.  It allows us to use a canonical formalism, hence
gaining  intuition by comparison with well studied systems in
non-relativistic quantum statistical mechanics. Our results are still
covariantly, though not manifestly, since we have restricted ourselves
to an invariant choice of coarse-graining.\\
2.  To perform a  perturbation expansion for the equations of motion
it is sufficent to construct perturbatively the field
propagator $e^{-i \hat{H}t}$. This is best carried out in the Fock
representation \cite{Ber}, which turns out to be particularly useful
for implementing our choice of coarse graining. \\
3.  We are  allowed  a certain degree of flexibility since the choice of the
projector onto the level of description is not unique
\cite{RaMi}. Hence, depending on the details of our problem (mainly
the initial condition) we can choose a projector so as to reduce the
strength of the noise terms. \\
4.  It provides a straightforward relation between the initial state
of the irrelevant variables and the noise terms in the evolution
equations. \\
5. It does not depend on the particular dynamics of the full system,
that is one can apply it even when the field evolution is
non-unitary, non-Markovian or non-autonomous . Therefore, it might be used in
conjunction with other 
methods (in particular the influence functional technique) in order to
reduce the amount of calculations needed for a particular problem. \\
6.  The Zwanzig method, is essentially algebraic, in the sense that
it depends 
solely on the properties of the space of observables and not on any
particular realization in some Hilbert space. This means that, at
least in principle, one can employ it in systems where quantum are
coupled to classical variables as is the case of the field theory in
curved spacetime. 
\paragraph{This paper}
It is the aim of this paper to apply the above ideas in the simplest
of field systems, first to exhibit the technique and understand the
insight it can offer in particular for the case of quantum to
classical transition. Hence, we concentrate on a single
self-interacting scalar field in Minkowski spacetime and consider
 coarse-grainings corresponding to truncation
at the levels $n=1$ and $n=2$. 
\par
We mainly focus on two issues:  the
derivation of the effectiver equation for the relevant variables and
the estimation of strength of the noise term, which determines the degree of
predictability of our preferred set of variables.
\par
The paper is then organized as follows: In section 2 we give a brief
review of the Zwanzig projection formalism, and construct the
indempotent operators that implement the coarse-graining operation in
the space of observables. In section 3, we derive the mean field
dynamics ina $\lambda \phi^4$ scalar field theory and give a general
discussion on the relation of correlation noise with the initial
condition. In section 4 we perform the same analysis for a $g \phi^3$
theory for the case of truncation at the level of two point functions. 
Finally in section 5, we give a discussion of our results, on the
possibility of obtaining Markovin behaviour and on future applications
of the formalism.
\par
We have found more convenient to implement the coarse-graining on the
normal-ordered form of the observables. The expressions we obtain are
simplified significantly if we use an index notation to denote
products of creation and annihilation operators $\hat{a}({\bf x})$ and
$\hat{a}^{\dagger}({\bf x}) $. The conventions of this notation are found in
Appendix 1. Finally, some useful formulas concerning the Fock
representation and the normal form of operators are to be found in
Appendix 2.

   \section{The method}

\subsection{The Zwanzig technique}
We will give a brief summary of the Zwanzig projection formalism, 
following the conventions of Zeh \cite{Zeh}.
The main idea in the Zwanzig formalism is the representation of the
coarse graining operator by an indempotent mapping ${\bf P}$ in the
space of states
\begin{equation}
\rho \rightarrow \rho_{rel} =  {\bf P} \rho \hspace{2cm} {\bf P}^2 =
{\bf P}
\end{equation}
The irrelevant part of the state is then given by 
\begin{equation}
\rho_{irr} = ({\bf 1} - {\bf P}) \rho
\end{equation}
${\bf P}$ is essentially a projection operator in the space of states
and determines through the trace functional a conjugate projector
${\bf P}^*$ on 
the space of observables. The projector needs not be self-adjoint
 ( ${\bf P} = {\bf P}^*$) but for convenience we shall assume so. 
\par
The projector {\bf P} determine the level of description for our
system . We should remark, that the choice of ${\bf P}$ projecting to
a particular class of observables is not unique; there can be
different inequivalent choices. Strictly speaking, ${\bf P}$ should be
considered as an operation on the states of the system and only in
this sense is it unique.
\par
To obtain the evolution equation for the relevant observables one
starts from the full dynamics of our system. The formalism is not
restricted to unitary dynamics; it can be applied equally well when
the dynamics are non-unitary or non-Markovian or non-local in
time. In our case, we shall restrict ourselves to unitary evolution
given through the von-Neumann equation
\begin{equation}
i \frac{\partial \rho}{\partial t} = {\bf L} \rho \equiv [H,\rho]
\end{equation}  
from which we obtain the following system of coupled differential equations for
$\rho_{rel}$ and $\rho_{irr}$
\begin{eqnarray}
i \frac{\partial \rho_{rel}}{ \partial t} &=& {\bf PL} \rho_{rel} +
{\bf PL} \rho_{irr} \\
i \frac{\partial \rho_{irr}}{ \partial t} &=& ({\bf 1} -  {\bf P}){\bf
L} \rho_{rel} +
({\bf 1}- {\bf P}){\bf L} \rho_{irr}
\end{eqnarray}
We can solve equation (2.5) by treationg the $\rho_{rel}$ term as an
external force
\begin{equation}
\rho_{irr}(t) = e^{-i ({\bf 1} - {\bf P}) {\bf L}t} \rho_{irr}(0) - i
\int_0^t d \tau e^{-i ({\bf 1} - {\bf P}) {\bf L} \tau} ({\bf 1} - {\bf
P}) {\bf L} \rho_{rel}(t - \tau)
\end{equation}
Here we have denoted by $e^{-i ({\bf 1} - {\bf P}) {\bf L}t} = 
({\bf 1} - {\bf P}) e^{-i {\bf L} t}$ the
evolution operator of the equation
\begin{equation}
i \frac{\partial \rho}{\partial t} = ({\bf 1} - {\bf P}) {\bf L} \rho
\end{equation}
and no actual exponentiation is implied. 
Substituting (2.6) into (2.4) we get the Zwanzig pre-master equation
\begin{eqnarray}
i \frac{\partial \rho_{rel}(t)}{ \partial t} = {\bf PL} \rho_{rel}(t)
+ {\bf PL}  e^{-i ({\bf 1} - {\bf P}) {\bf L}t} \rho_{irr}(0) -i
\int_0^t d \tau {\bf G}(\tau) \rho_{rel}(t-\tau)
\end{eqnarray}
Here ${\bf G}$ stands for the kernel 
\begin{equation}
{\bf G}(\tau) = {\bf PL}e^{-i ({\bf 1} - {\bf P}) {\bf L} \tau} ({\bf 1} -
{\bf P}) {\bf LP}
\end{equation}
Given then a relevant observable $A$, i.e. one such that ${\bf P} A =
A$, we obtain for the evolution of its expectation value $ \langle
A \rangle$
\begin{eqnarray}
i \frac{\partial}{\partial t} \langle A \rangle(t) -  \langle
{\bf PL} A \rangle(t) \hspace{3cm} \nonumber \\
+ i \int_0^t d \tau \langle {\bf PL}({\bf
1}- {\bf P}) e^{i  {\bf L} \tau} ({\bf
1}- {\bf P}) {\bf LP} A \rangle(t-\tau) = F_A(t)
\end{eqnarray}
$F_A(t)$ is ``driving force'' term, essentially stochastic in nature, 
since it
depends on the irrelevant components of the initial state that are
inaccesible from our level of description. It reads
\begin{equation}
 F_A(t) = - Tr \left( \rho(0) [({\bf 1} 
- {\bf P}) e^{i ({\bf 1} 
- {\bf P}) {\bf L} t} {\bf L} A] \right)
\end{equation}
\par
Note, that in general the evolution of the relevant observables is
non-local in time.

\subsection{The coarse-graining operator}
Equation (2.11) provides is the starting point for a detailed
calculation of the evolution equations for  the relevant
observables. The only input one needs 
give is the particular form of the coarse-graining operator ${\bf
P}$. 
\par
We want ${\bf P}$ to correspond as close as possible to the notion of
the truncation of the hierarchy of  correlation functions at some
order $n$. To see how one can proceed in the construction, let us
examine first the case for $n=1$. Here, the relevant variables are the
values of the field $\hat{\phi}({\bf x})$ at each given instant of
time. Recall, that the field operator can be written in terms of
creation and annihilation operators. Then consider any density matrix
writen in normal-ordered form
\begin{equation}
\rho = \sum_{r,s} \hat{a}^{\dagger}_{a_1} \ldots \hat{a}^{\dagger}_{a_r}
\rho^{a_1 \ldots a_r}_{\quad \quad b_1 \ldots b_s} \hat{a}^{b_1} \ldots
\hat{a}^{b_s} 
\end{equation}
We remark that the contributions to the expectation value of $\phi$
arise solely from the terms in the summation characterized by $r = s +
1$ or $r = s - 1$. That is, only terms differing in the number of
$\hat{a}$ 's and $a^{\dagger}$ 's by one are the contributing ones. 
\par
 Requiring thet  ${\bf P}$ projects any operator into a linear
combination of $\hat{a}$ 's and $a^{\dagger}$ ' (this corresponds to
considering field and momentum expectation values for relevant
observables as is natural in a canonical treatment) and taking the
above remark into consideration, we arrive at a natural choice for the
projector. Write any observable into its normal ordered form 
\begin{equation}
\hat{A} = \sum_{r,s} \hat{a}^{\dagger}_{a_1} \ldots \hat{a}^{\dagger}_{a_r}
A^{a_1 \ldots a_r}_{\quad \quad b_1 \ldots b_s} \hat{a}^{b_1} \ldots
\hat{a}^{b_s} 
\end{equation}
and implement the action of ${\bf P}$ in each term in the series as
follows: if $|r = s| \neq 1$ then the action of ${\bf P}$ yields
zero. If $r = s + 1 $ then 
\begin{equation}
{\bf P} \left( \hat{a}^{\dagger}_{a_1} \ldots \hat{a}^{\dagger}_{a_s+1}
A^{a_1 \ldots a_{s+1}}_{\;\;\quad \quad b_1 \ldots b_s} \hat{a}^{b_1} \ldots
\hat{a}^{b_s}\right) = \hat{a}^{\dagger}_a K^a
\end{equation}
with $K^a$ is obtained by summing over all possible contractions of
the $s+1$ upper indices with the $s$ lower ones. 
\par
Let us give one
simple example to illustrate this. Consider a term of the form
$A = \hat{a}^{\dagger}_a \hat{a}^{\dagger}_b A^{ab}_{\;\; \;c}
\hat{a}^c$. 
The action
of ${\bf P}$ reads
\begin{equation}
{\bf P} \hat{A} = \hat{a}^{\dagger}_a A^{ab}_{\;\;\;b} + \hat{a}^{\dagger}_b
A^{ab}_{\;\;\;a}  
\end{equation}
We proceed similarly for the case $r = s - 1$.
\par
The generalization for higher order products of operators follows
along the same lines. Consider for instance a level of description
fixed at one- and two- point correlation functions. We then have ${\bf P}$
projecting onto linear combinations of operators of the form 
$\hat{a}$, $\hat{a}^{\dagger}$, $\hat{a} \hat{a}$, $\hat{a}^{\dagger}
\hat{a}$ and $\hat{a}^{\dagger} \hat{a}^{\dagger}$. When acting on any
normal-ordered operator ${\bf P}$ will yield a non-zero expression if
$ |r - s| \in \{ 0, 1, 2 \}$. For example, consider a term
$\hat{a}^{\dagger}_a \hat{a}^{\dagger}_a A^{ab}_{\;\;\;cd} \hat{a}^c
\hat{a}^d $. Action with ${\bf P}$ will yield
\begin{equation}
\hat{a}^{\dagger}_a A^{ab}_{\;\;\;cb} \hat{a}^c + 
\hat{a}^{\dagger}_a A^{ab}_{\;\;\;bd} \hat{a}^d + 
\hat{a}^{\dagger}_b A^{ab}_{\;\;\;ca} \hat{a}^c + 
\hat{a}^{\dagger}_a A^{ab}_{\;\;\;ad} \hat{a}^d 
\end{equation} 

\subsection{Perturbation expansion}
Having identified ${\bf P}$ we are only left with the calculation of
the terms appearing in equation (2.11). In the following we shall
assume that the Hamiltonian is of the form $\hat{H} = \hat{H}_0 +
\hat{V}$. We should note that evolution according to the free
Hamiltonian does not change the level of description (since $ {\bf
L_0 P} = {\bf P L_0}$ where ${\bf L}_0 \rho = [\hat{H}_0,\rho]$) and
therefore the expression of the non-local term simplifies
\begin{equation}
 i \int_0^t d \tau \langle {\bf PV}({\bf
1}- {\bf P}) e^{i  {\bf L} \tau} ({\bf
1}- {\bf P}) {\bf VP} A \rangle(t-\tau) 
\end{equation}
where ${\bf V}\rho = [\hat{V},\rho]$.
From this expression we can readily see that in a perturbative
expansion the local in time term will be at least of second order to
the coupling constant. This is easily understood since this term
comes from 
correlations, that start as relevant at time $0$, become irrelevant due
to interaction at time $\tau$,  propagate as irrelevant and become 
relevant again at time $t$. Hence in the perturbative expansion at least
diagrams containing two vertices are having non-zero contribution. On the other
hand, the noise term, containing the evolution of correlations
starting and propagating as irrelevant and due to an interaction at
time $t$ becoming relevant, can be of the first order to the coupling
constant thus being dominant in lowest part of the perturbation
series.
This means that unless we consider some
particular initial condition the effect of the noise might destroy any
sense of predictability for our selected variables. 
\par
Another important observation is that the potential appears in the
non-local term only in the combination ${\bf PV}$. This part of the
potential essentially scatters relevant information 
only to a particular sector of irrelevant states ( these are sometimes
called ``doorway states'' \cite{Zeh}). For example, in the $g \phi^3$
theory with truncation at the level of $n=2$, we shall examine in the
following sections, the doorway states are the ones supporting
 third order correlations. Further propagation is needed to
reach states with higher order correlations.
\par
When considering the lowest order term in the perturbation expansion
the expression of the non-local terms is significantly simplified.
To see this, note that these can be writen in the form 
\begin{equation}
 i \int_0^t d \tau \left( {\bf P} [\hat{V},   ({\bf 1} - {\bf P}) \left( e^{-i\hat{H}_0\tau}[\hat{V}, 
\rho_{rel}(t-\tau)] e^{i\hat{H}_0 \tau}\right)], A \right)
\end{equation}
where $(,)$ refers to the Hilbert-Schmidt inner product. Now since
$ || e^{-iH_0 \tau} \rho(t-\tau) e^{iH_0 \tau} - \rho(t) || =
O(g) $
we can easily verify that within the second order to the coupling
constant we get
\begin{equation}
i \langle    {\bf P} \left(             [  ({\bf 1} - {\bf P}) \left([\hat{A}, \hat{V}]
\right), W] \right) \rangle(t)
\end{equation}
where
\begin{equation}
\hat{W}(t) = \int_0^t d \tau e^{- i \hat{H}_0 \tau} \hat{V} e^{i
\hat{H}_0 \tau} 
\end{equation}
Hence to the lowest order in the perturbative expansion the
non-unitary term becomes local in time. This is due to the fact
that the free propagation can  not remove
correlations from the doorway states into the more deeply lying states
of the irrelevant sector. Evolution within the sector of doorway
states makes the correlations lose fast the memory of the initial
condition (within a time interval proportional to the coupling constant) and
hence when they reappear in the relevant channel they do not impose a
time correlation in the relevant dynamics.  
\par
We are going to carry our calculation in the lowest order of
perturbation theory. We should remark though, that apart from the
technical complication, the computation of higher order corrections is
not difficult. It is sufficient to have a perturbation expansion in
the propagator $e^{-i\hat{H}t}$. This is best carried in the Fock
representation \cite{Ber}, which is a desirable feature given the
connection of our coarse-graining projector with the normal-ordered
form of the observables.

\section{Evolution of mean field in $\lambda \phi^4 $ theory}

Let us apply now the above construction to the case of a $\lambda
\phi^4 $ theory for truncation at the level $n=1$. 
The operator for the potential is given by equations
(B.14- B.19), while the operator $\hat{W}$ is easily computed
\begin{eqnarray}
\hat{W} = \frac{\lambda}{4!} \left( W_{abcd} \hat{a}^a \hat{a}^b
\hat{a}^c \hat{a}^d + 4 \hat{a}^{\dagger}_a W^a_{\;\;bcd}\hat{a}^b
\hat{a}^c \hat{a}^d \right. \nonumber \\ 
\left. + 6 \hat{a}^{\dagger}_a \hat{a}^{\dagger}_b 
W^{ab}_{\quad cd} \hat{a}^c \hat{a}^d + 4  \hat{a}^{\dagger}_a
\hat{a}^{\dagger}_b  \hat{a}^{\dagger}_c W^{abc}_{\quad \; d}
\hat{a}^d + \hat{a}^{\dagger}_a \hat{a}^{\dagger}_b
\hat{a}^{\dagger}_c \hat{a}^{\dagger}_d W^{abcd}\right)
\end{eqnarray}
with
\begin{eqnarray}
W_{abcd} \leadsto  \int \prod_{i=1}^{4} \frac{dk_i}{(2
\omega_{{\bf k}_i})^{1/2} } e^{-i({\bf k}_1{\bf x}_1+{\bf k}_2 {\bf x}_2 +{\bf k}_3 {\bf x}_3+{\bf k}_4{\bf x}_4)} \; (2
\pi)^3 \delta({\bf k}_1+{\bf k}_2 +{\bf k}_3 +{\bf k}_4)
\hspace{1cm} \nonumber \\
 \times
\frac{ e^{-i(\omega_{{\bf k}_1} + \omega_{{\bf k}_2} +
\omega_{{\bf k}_3}+ \omega_{{\bf k}_4})t} - 1}{-i(\omega_{{\bf k}_1} + \omega_{{\bf k}_2} +
\omega_{{\bf k}_3}) +\omega_{{\bf k}_4}}
\hspace{2cm}
\\
W^a_{\;\;bcd} \leadsto  \int \prod_{i=1}^{4} \frac{dk_i}{(2
\omega_{{\bf k}_i})^{1/2} } e^{-i(-{\bf k}_1{\bf x}_1+{\bf k}_2 {\bf x}_2 +{\bf k}_3 {\bf x}_3+{\bf k}_4{\bf x}_4)} \; (2
\pi)^3 \delta({\bf k}_1+{\bf k}_2 +{\bf k}_3 +{\bf k}_4)
\hspace{1cm} \nonumber \\
 \times
\frac{e^{-i(-\omega_{{\bf k}_1} + \omega_{{\bf k}_2} +
\omega_{{\bf k}_3}+ \omega_{{\bf k}_4})t}- 1}{-i(-\omega_{{\bf k}_1} + \omega_{{\bf k}_2} +
\omega_{{\bf k}_3}) +\omega_{{\bf k}_4}}
\hspace{2cm}
\\ 
W^{ab}_{\quad cd} \leadsto  \int \prod_{i=1}^{4} \frac{dk_i}{(2
\omega_{{\bf k}_i})^{1/2} } e^{-i(-{\bf k}_1{\bf x}_1-{\bf k}_2 {\bf x}_2 +{\bf k}_3 {\bf x}_3+{\bf k}_4{\bf x}_4)} \; (2
\pi)^3 \delta({\bf k}_1+{\bf k}_2 +{\bf k}_3 +{\bf k}_4)
\hspace{1cm} \nonumber \\
 \times
\frac{ e^{-i(-\omega_{{\bf k}_1} - \omega_{{\bf k}_2} +
\omega_{{\bf k}_3}+ \omega_{{\bf k}_4})t} - 1 }{-i(-\omega_{{\bf k}_1} - \omega_{{\bf k}_2} +
\omega_{{\bf k}_3}) +\omega_{{\bf k}_4}}
\hspace{2cm}
\\
W^{abc}_{\quad \;\;d} \leadsto  \int \prod_{i=1}^{4} \frac{dk_i}{(2
\omega_{{\bf k}_i})^{1/2} } e^{-i(-{\bf k}_1{\bf x}_1-{\bf k}_2 {\bf x}_2 -{\bf k}_3 {\bf x}_3+{\bf k}_4{\bf x}_4)} \; (2
\pi)^3 \delta({\bf k}_1+{\bf k}_2 +{\bf k}_3 +{\bf k}_4)
\hspace{1cm} \nonumber \\
 \times
\frac{ e^{-i(-\omega_{{\bf k}_1} - \omega_{{\bf k}_2} -
\omega_{{\bf k}_3}+ \omega_{{\bf k}_4})t} - 1}{-i(-\omega_{{\bf k}_1} - \omega_{{\bf k}_2} -
\omega_{{\bf k}_3}) +\omega_{{\bf k}_4}}
\hspace{2cm}
\\
W^{abcd} \leadsto  \int \prod_{i=1}^{4} \frac{dk_i}{(2
\omega_{{\bf k}_i})^{1/2} } e^{i({\bf k}_1{\bf x}_1+{\bf k}_2 {\bf x}_2 +{\bf k}_3 {\bf x}_3+{\bf k}_4{\bf x}_4)} \; (2
\pi)^3 \delta({\bf k}_1+{\bf k}_2 +{\bf k}_3 +{\bf k}_4)
\hspace{1cm} \nonumber \\
 \times
\frac{ e^{i(\omega_{{\bf k}_1} + \omega_{{\bf k}_2} +
\omega_{{\bf k}_3}+ \omega_{{\bf k}_4})t} - 1}{i(\omega_{{\bf k}_1} + \omega_{{\bf k}_2} +
\omega_{{\bf k}_3}) +\omega_{{\bf k}_4}}
\hspace{2cm}
\end{eqnarray}
\par
Having the expression for $\hat{W}$  one can use in a straightforward
way equation (2.19) to 
compute the dissipative terms in the evolution equation. Let us perform the
calculations step by step.
\\ 
First we compute the commutator $[\hat{a}^a,\hat{V}]$. It reads
\begin{equation}
[\hat{a}^a,\hat{V}] = \frac{\lambda}{4!} \left( 4 V^a_{\;\;bcd}
\hat{a}^b \hat{a}^c \hat{a}^d + 12 \hat{a}^{\dagger}_b V^{ab}_{\quad
cd}  \hat{a}^c \hat{a}^d + 12  \hat{a}^{\dagger}_b \hat{a}^{\dagger}_c
V^{abc}_{\quad \;d} \hat{a}^d + \hat{a}^{\dagger}_b
\hat{a}^{\dagger}_c \hat{a}^{\dagger}_d V^{abcd} \right)
\end{equation}
Acting the projector ${\bf P}$ on this we obtain
\begin{equation}
{\bf P}  [\hat{a}^a,\hat{V}] = \frac{\lambda}{4!} \left( 24
V^{ac}_{\quad cb} \hat{a}^c + 24 \hat{a}^{\dagger}_b V^{abc}_{\quad c} \right)
\end{equation}
Hence we can easily read the operator $( {\bf 1} - {\bf P} )
[\hat{a}^a,\hat{V}] $. 
\par
One, then, needs to compute its commutator with the operator
$\hat{W}$. This is indeed the difficult part of the calculations. We
will get 24 terms, out of which only 12 will survive after the action
on them of ${\bf P}$. There is no need to reproduce the whole of the
calculations here, but for purposes of exposition we shall present the
computations involved in one term. 
\paragraph{An example}
We consider the term 
\begin{equation}
16 W^e_{\;\;fgh} V^{abcd} [ \hat{a}^{\dagger}_b  \hat{a}^{\dagger}_c
\hat{a}^{\dagger}_d, \hat{a}^{\dagger}_e \hat{a}^f \hat{a}^g
\hat{a}^h]  \nonumber
\end{equation} 
After computing the commutator we will obtain
\begin{equation}
- 16 \left[ 9 \hat{a}^{\dagger}_e \hat{a}^{\dagger}_c 
 \hat{a}^{\dagger}_d 
W^e_{\;\;fgb} V^{abcd} \hat{a}^f \hat{a}^g + 18 \hat{a}^{\dagger}_e
\hat{a}^{\dagger}_f   W^e_{\;\; fbc} V^{abcd} \hat{a}^f + 6
\hat{a}^{\dagger}_e W^{e}_{\;\; bcd} V^{abcd} \right]
\end{equation}
The action of ${\bf P}$ on  (3.10) will yield
\begin{eqnarray}
-16 \left[ 9 \, (4 \hat{a}^{\dagger}_c W^e_{\;\;ebd}V^{abcd} + 2 \hat{a}^{\dagger}_e
W^e_{\;\;bcd} V^{abcd}) + 12 \hat{a}^{\dagger}_d W^e_{\;\;ebc}
V^{abcd} \right. \nonumber \\
\left. + 12 \hat{a}^{\dagger}_e W^e_{\;\;bcd} V^{abcd} + 6
\hat{a}^{\dagger}_e W^e_{\;\;bcd} V^{abcd} \right) \nonumber \\
= -16 \cdot 12 \left( 4 \hat{a}^{\dagger}_d W^e_{\;\;ebc} + 3
\hat{a}^{\dagger}_e W^e_{\;\;bcd} V^{abcd} \right)
\end{eqnarray}
\paragraph{The evolution equations}
The final result reads
\begin{eqnarray}
\lambda^2 \left[ \frac{3}{2} (W^{cbd}_{\quad \;\; d} V^a_{\;\;bce} -
W^d_{\;\;dbc}V^{abc}_{\quad \;\;e} \hspace{3cm} \right.
\nonumber \\
\left. +( - W_{bcde}V^{abcd} - W^d_{\;\;ebc} V^{abc}_{\quad \;\; d} +
W^{cd}_{\quad be}V^{ab}_{\quad cd} + W^{bcd}_{\quad \;\; e}
V^a_{\;\;bcd}) \right] \nonumber \\
+ \lambda^2 \hat{a}^{\dagger}_e \left[ \frac{3}{2} ( W^{cbd}_{\quad \;\;d} V^{ae}_{\quad
cb} - W^d_{\;\;dbc} V^{abce}) \hspace{3cm} \right.
\nonumber \\
\left. +(W^{bcde}V^a_{\;\;bcd} + W^{cde}_{\quad \;\;b} V^{ab}_{\quad
cd} - W^{de}_{\quad bc} V^{abc}_{\quad \;\;d} - W^e_{\;\;bcd} V^{abcd}
) \right]
\end{eqnarray}
Note, the symmetry between the terms contracting $\hat{a}^e$ and
$\hat{a}^{\dagger}_e$. 
\par
We can therefore write down the evolution equation for $a({\bf x}) = \langle
\hat{a}({\bf x}) \rangle $ and $a^*({\bf x}) = \langle \hat{a}^{\dagger}({\bf x})
\rangle $:
\begin{eqnarray}
i \frac{\partial}{\partial t} a({\bf x}) - \int d{\bf x}' h({\bf x},{\bf x}') a({\bf x}') - \lambda
\int d x  \left( V({\bf x},{\bf x}')a({\bf x}') + V({\bf x},-{\bf x}') a^*({\bf x}') \right) \nonumber 
\\
- i \lambda^2  \int dx' \left( A({\bf x},{\bf x}') a({\bf x}') + A^*({\bf x},-{\bf x}') a^*({\bf x}')\right) \hspace{3cm}
\nonumber \\
+  \lambda^2 \int dx' \left( B({\bf x},{\bf x}') a({\bf x}') + B({\bf x},-{\bf x}') a^*({\bf x}') \right)
= F_{a({\bf x})}(t) \hspace{2cm}
\end{eqnarray}
where  h({\bf x},{\bf x}') is given by equation (B.2), $V({\bf x},{\bf x}')$  (essentially
$V^{ac}_{\quad cb}$) reads
\begin{equation}
V({\bf x},{\bf x}') = \int \frac{dk_1}{(2 \omega_{{\bf k}_1})^{1/2}}  \frac{d{\bf k}_2}{(2
\omega_{{\bf k}_2})^{1/2}}   \frac{1}{2 \omega_{({\bf k}_1+{\bf k}_2)/2}} e^{-i{\bf k}_1{\bf x} +
i{\bf k}_2 {\bf x}'}
\end{equation}
while 
\begin{eqnarray}
A({\bf x},{\bf x}') = \int \prod_{i=1}^{4} \frac{d k_i}{2
\omega_{{\bf k}_i} } e^{-i{\bf k}_1({\bf x}-{\bf x}')} \; (2
\pi)^3 \delta({\bf k}_1+{\bf k}_2 +{\bf k}_3 +{\bf k}_4) \Delta({\bf k}_1,{\bf k}_2,{\bf k}_3,{\bf k}_4;t)\hspace{1cm}
\\
B({\bf x},{\bf x}') = 3 \; \int \frac{dk_1}{(2 \omega_{{\bf k}_1})^{1/2}} \frac{d{\bf k}_2}{(2
\omega_{{\bf k}_2})^{1/2}} e^{i{\bf k}_1{\bf x}-i{\bf k}_2{\bf x}'} \hspace{3cm}\nonumber \\
\times 
\left( \int \frac{d{\bf k}_1}{2
\omega_{{\bf k}_3}} \frac{d{\bf k}_1}{2 
\omega_{{\bf k}_4}} \; (2 \pi)^3 \delta({\bf k}_1+{\bf k}_2+{\bf k}_3+{\bf k}_4) E({\bf k}_3,{\bf k}_4;t)
\right) \hspace{2cm}
\end{eqnarray}
\par
$\Delta$ and $E$ contain the time dependence of the kernels $A$ and
$B$ and read
\begin{eqnarray}
\Delta ({\bf k}_1,{\bf k}_2,{\bf k}_3,{\bf k}_4;t) =  \int_0^t d \tau e^{-i \omega_{{\bf k}_1}\tau}
\left (- e^{-i(\omega_{{\bf k}_2}+\omega_{{\bf k}_3} +\omega_{{\bf k}_4})\tau}
\right. \nonumber \\
\left.
- e^{-i(\omega_{{\bf k}_2}+\omega_{{\bf k}_3} - \omega_{{\bf k}_4})\tau}
+ e^{-i(\omega_{{\bf k}_2}-\omega_{{\bf k}_3} -\omega_{{\bf k}_4})\tau}
+ e^{i(\omega_{{\bf k}_2}+\omega_{{\bf k}_3} +\omega_{{\bf k}_4})\tau} \right)
\\
E({\bf k},{\bf k}';t) =  \frac{ \cos(\omega_{{\bf k}}+\omega_{{\bf k}'})t - 1}{\omega_{\bf k} +\omega_{{\bf k}'}}
\end{eqnarray}
Note that for times $t << m^{-1}$ we have $\Delta({\bf k}_1,{\bf k}_2,{\bf k}_3,{\bf k}_4;t)
\approx t$.
\par
A more transparent form is given when calculating the expectation
values of creation and annihilation operators in momentum space.
\begin{eqnarray}
\frac{\partial}{\partial t} a({\bf k}) + i \omega_{\bf k} a({\bf k}) + i \int dk'
\left[ V({\bf k},{\bf k}') - \lambda B({\bf k},{\bf k}') \right] [ a({\bf k}') +a^*({\bf k}') ]
\nonumber \\  \\
- \lambda^2 \left[ A({\bf k};t) a({\bf k})+ A^*({\bf k};t) a^*({\bf k})\right] = -i F_{a({\bf k})}(t) \hspace{2cm} \\ 
\frac{\partial}{\partial t} a^*({\bf k})  -i \omega_{\bf k} a^*({\bf k}) - i \int dk'
\left[ V({\bf k},{\bf k}') - \lambda B({\bf k},{\bf k}') \right][ a({\bf k}') +a^*({\bf k}') ]
\nonumber \\ 
- \lambda^2 \left[ A({\bf k},t) a({\bf k}) +A^*({\bf k};t) a^*({\bf k}) \right] 
  = - i F_{a^*({\bf k})}(t) \hspace{2cm}
\end{eqnarray}  
with
\begin{eqnarray}
A({\bf k}) = \frac{1}{2 \omega_{\bf k}} \int \frac{dk_1}{2 \omega_{{\bf k}_1}}
\frac{dk_2}{2 \omega_{{\bf k}_2}} \frac{1}{2 \omega_{{\bf k}+{\bf k}_1+{\bf k}_2}}
\Delta({\bf k},{\bf k}_1,{\bf k}_2,{\bf k}+{\bf k}_1+{\bf k}_2;t) \\
B({\bf k},{\bf k}') = \frac{3}{16 \omega_{{\bf k}+{\bf k}'}\omega_{{\bf k}'}} 
\int \frac{dk_1}{ \omega_{{\bf k}_1} \omega_{{\bf k}+{\bf k}'+{\bf k}_1}} E({\bf k}_1,{\bf k}+{\bf k}'+{\bf k}_1;t) \\
V({\bf k},{\bf k}') = \frac{1}{4  \omega_{{\bf k}'} \omega_{({\bf k}+{\bf k}')/2}} \hspace{4cm}
\end{eqnarray}
\paragraph{Renormalization}
The function $A({\bf k};t)$ is actually divergent. We can perform a Taylor
expansion of $A$ around ${\bf k} = 0$ and verify that the term $A(0;t)$ is
divergent, the terms containing first derivatives vanish while the
ones containing the second order derivatives are finite. Hence, as
could be expected, it is the zero modes of the field that give a
divergent contribution. This can be removed by a redefinition
\begin{equation}
A_{ren}({\bf k};t) = A({\bf k};t) - A(0;t)
\end{equation}
and by absorbing $A(0;t)$ in a field renormalization. To see this,
note that
\begin{equation}
\frac{\partial}{\partial t} \left( \begin{array}{c} 
                                    a({\bf k}) \\
                                    a^*({\bf k})
                                    \end{array} \right) = 
\mbox{\cal finite terms} + \lambda^2 \left( \begin{array}{cc}
                           A(0;t)  &   A^*(0;t) \\ 
                           A(0;t)  &   A^*(0;t) \\
                           \end{array} \right)   
\                               \;\;   \left( \begin{array}{c} 
                                     a({\bf k}) \\
                                     a^*({\bf k})
                                     \end{array} \right)
\end{equation}
Hence the divergencies can be absorbed through a redefinition of the
Heisenberg picture operators $\hat{a}({\bf k},t)$,  $\hat{a}^{\dagger}({\bf k},t)$ 
\begin{equation}
     \left( \begin{array}{c} 
     \hat{a}({\bf k},t) \\
     \hat{a}^{\dagger}({\bf k},t)
     \end{array} \right)          \rightarrow \exp \left[ \lambda^2 \int_0^t d
\tau \left( \begin{array}{cc}
                           A(0;t)  &   A^*(0;t) \\ 
                           A(0;t)  &   A^*(0;t) \\
                           \end{array} \right)      \right] 
                                                \left( \begin{array}{c} 
     \hat{a}({\bf k},t) \\
     \hat{a}^{\dagger}({\bf k},t)
     \end{array} \right)     
\end{equation}
It is easy to interpret the terms in (3.19) and (3.20). The term $V$
contains the lowest order contribution from the potential to our
coarse-grained dynamics. Its form is better understood by observing
that the mean field theory approximation amounts to substituting
four-point vartices (say with incoming momenta ${\bf k}_1$ and ${\bf k}_2$   
and outcoming ${\bf k}_3$ and ${\bf k}_4$) with free propagation of a mode with
momentum the average of the incoming (or the outcoming) modes'
momenta: $({\bf k}_1+{\bf k}_2)/2$. 
\par
The term $B$ is a higher order, time dependent correction to the
contribution of the potential, while the term $A$ corresponds to
dissipation. This is easily verified when we take the time-reverse of
equations (3.19) and (3.20). The terms containing $A$ are the only
non-invariant terms.

\paragraph{The noise terms}
Most important, from the point of view of the classical behaviour and
predictability of the mean field is the noise term. As we said it is
at least of first order to the coupling constant and in principle can
dominate  both the potential and the dissipation terms.
\par
Starting from equation (2.11) it is straightforward to calculate the
leading (first order to $\lambda$) contribution to the noise. It reads
(we switch back to the index notation)
\begin{equation}
F_{a^a}(t) = \frac{\lambda}{4!} \; Tr \left( \rho(0) \hat{A}(t) \right)
\end{equation}
where 
\begin{eqnarray}
\hat{A}(t) = 4 V^a_{\;\;bcd} \hat{a}^b(t) \hat{a}^c(t)\hat{a}^d(t)
+ 12 \hat{a}^{\dagger}_b(t) V^{ab}_{\quad cd} \hat{a}^c(t)\hat{a}^d(t) 
\nonumber \\
+12 \hat{a}^{\dagger}_b(t)\hat{a}^{\dagger}_c(t) V^{abc}_{\quad\; d}
\hat{a}^d(t) + 4
\hat{a}^{\dagger}_b(t)\hat{a}^{\dagger}_c(t)\hat{a}^{\dagger}_d(t)
V^{abcd} \nonumber \\
-24 V^{ac}_{\quad \;cb} \hat{a}^b(t) -24 \hat{a}^{\dagger}_b
V^{abc}_{\quad \;\;c} \hspace{2cm}
\end{eqnarray}
where with $\hat{a}(t)$ and $\hat{a}^{\dagger}(t)$ we denote the
Heisenberg picture operators evolving according to the free Hamiltonian.
\par
In order for our coarse-grained description to satisfy the
predictability criterion, the noise term should be sufficiently weak
(though strong enough to cause decoherence of the mean field
histories). This, as we see, cannot be true for a generic initial
state of the system. We can nevertheless observe that the noise terms
vanishes when the initial state is the vacuum $\rho_{vac} = |0 \rangle
\langle 0|$. This means that for states $\rho(0)$ sufficiently close
to the vacuum the noise term becomes smaller and smaller. This means
that for any state $\rho(0)$ such that $|| \rho(0) - \rho_{vac}||_{HS} <
\epsilon$ the noise term will be of order $O(\epsilon)$.
\par
Consider for instance that the initial state of the system is some
coherent state $|\alpha({\bf x}) \rangle$, determined by a square-integrable
function $\alpha({\bf x})$. Coherent states are eigenstates of the
annihilation operators, hence the trace in equation (3.) is easily performed.
Now if we assume that $||\alpha({\bf x})|| < \epsilon$ it is easy to
establish that $|| \rho(0) - \rho_{vac}||_{HS} = O(\epsilon)$. Hence
to leading order in $\epsilon$ the noise term reads
\begin{equation}
F_{a^a}(t) = - \lambda \epsilon \left( V^{ac}_{\quad cb} \zeta^b(t) + 
\zeta^*_b(t) V^{abc}_{\quad \;\; c} \right)
\end{equation}
where we wrote $\alpha(t) = \epsilon \zeta(t)$. 
This is an example of an initial condition that renders the noise term
sufficiently weak to allow for predictability. This particular
condition, we believe, is  realistic when considering
cosmological scenaria.
\par
Finally we should remark that it is straightforward to obtain
evolution equations for the mean field and momentum by using the
equations
\begin{eqnarray}
\hat{a}({\bf k}) =  \int dx e^{i{\bf k}{\bf x}} \left( \omega_{\bf k} \hat{\phi}({\bf x}) + i
\hat{\pi}({\bf x}) \right) \\
\hat{a}^{\dagger}({\bf k}) =  \int dx e^{-i{\bf k}{\bf x}} \left( \omega_{\bf k} \hat{\phi}({\bf x}) - i
\hat{\pi}({\bf x}) \right) 
\end{eqnarray}
\section{Two point functions in $g \phi^3$ theory}

In this section we are going to give the results for the truncation of the
hierarchy at the level $n=2$ for a $g \phi^3$ scalar field theory.
\paragraph{ The mean field equations}
For completeness we will give very briefly the results of the mean field analysis for
the $g \phi^3$ case. The expectation value of the operator
$\hat{a}({\bf k})$ evolves according to an equation similar to (3.19)
\begin{eqnarray}
\frac{\partial}{\partial t} a({\bf k}) + i \omega_{\bf k} a({\bf k}) -i \lambda^2 \int
B({\bf k},{\bf k}') \left( a({\bf k}') + a^*({\bf k}') \right) \nonumber \\
- \lambda^2 \left[A({\bf k};t) a({\bf k}) + A^*({\bf k};t)a^*({\bf k}) \right] = F_{a({\bf k})}(t)
\end{eqnarray}
where here the functions $A$ and $B$ are given by
\begin{eqnarray}
A({\bf k};t) = \frac{3}{16 \omega_{\bf k}} \int \frac{dk_1}{\omega_{{\bf k}_3+{\bf k}}
\omega_{{\bf k}_3}} \Delta({\bf k},{\bf k}+{\bf k}_3,{\bf k}_3;t) \\
B({\bf k},{\bf k}';t) = \frac{1}{2 \omega_{{\bf k}'} \omega_{{\bf k}+{\bf k}'}} \frac{ \cos
\omega_{{\bf k}+{\bf k}'}t - 1  }{\omega_{{\bf k}+{\bf k}'}}
\end{eqnarray}
with
\begin{eqnarray}
   \Delta ({\bf k}_1,{\bf k}_2,{\bf k}_3;t) = - 
\; \;   2 i \int_0^t e^{-i\omega_{{\bf k}_1} \tau} \sin \left( \omega_{{\bf k}_2}
+ \omega_{{\bf k}_3} \right)
\end{eqnarray}
As is well known, the potential does not contribute in the lowest
order equation for the mean field theory, and the quantities $A$ and
$B$ again characterize dissipation and time-dependent correction to
the potential.
\paragraph{ Evolution equations for two-point functions}
Let us now give the results for the case of truncation at the $n =2 $
level. We prefer to give them in terms of the functions
$G({\bf k})$ and $Z({\bf k})$ defined by
\begin{eqnarray}
\langle \hat{a}({\bf x}) \hat{a}({\bf x}') \rangle = \int \frac{dk}{2 \omega_{\bf k}} e^{-i{\bf k}({\bf x}+{\bf x}')} G({\bf k}) \\
\langle \hat{a}^{\dagger}({\bf x}) \hat{a}({\bf x}') \rangle = \int \frac{dk}{2 \omega_{\bf k}}
e^{-i{\bf k}({\bf x}-{\bf x}')} Z({\bf k}) 
\end{eqnarray}
We will skip all calculations and present straightforwardly the
results, since the way to proceed is exactly as previously
 and the only difficulty
is a computational one. Thus, we get for a final result
\begin{eqnarray} 
\frac{\partial}{\partial t} a({\bf k}) + i \omega_{\bf k} a({\bf k}) = -i F_{a({\bf k})} \hspace{4cm}
\\ \\
\frac{\partial}{\partial t} Z({\bf k}) - \frac{\lambda}{4} \left(
\frac{2}{\omega_{\bf k}} +  \frac{1}{(\omega_{\bf k} \omega_{{\bf k}/2})^{1/2}} \right) [a({\bf k})
+ a^*({\bf k})] \hspace{2cm}
\nonumber \\ 
+i  \lambda^2 \int dk' \left[ r({\bf k},{\bf k}';t) G({\bf k}') + r^*({\bf k},{\bf k}';t) G^*({\bf k}') +
s({\bf k},{\bf k}';t) Z({\bf k}') \right] \nonumber \\
+i \lambda^2 \left[ D_1({\bf k};t) G({\bf k}) + D_2({\bf k};t) G^*({\bf k}) + D_3({\bf k};t) Z({\bf k}) \right]
= -i F_{Z({\bf k})}(t)
\\ \\
\frac{\partial}{\partial t} G({\bf k}) + 2 i \omega_{\bf k} G({\bf k})  - \frac{
\lambda}{4} \left( 
\frac{2}{\omega_{\bf k}} +  \frac{1}{(\omega_{\bf k} \omega_{{\bf k}/2})^{1/2}} \right) [a({\bf k})
+ a^*({\bf k})] \nonumber \\
+i  \lambda^2 \int dk' \left[  K_1({\bf k},{\bf k}';t) G({\bf k}') +  K_2({\bf k},{\bf k}';t) G^*({\bf k}') +
 K_3({\bf k},{\bf k}';t) Z({\bf k}') \right] \nonumber \\
+i \lambda^2 \left[ L_1({\bf k};t) G({\bf k}) + L_2({\bf k};t) Z({\bf k}) \right]
= -i F_{G({\bf k})}(t)
\end{eqnarray}
The form of the functions appearing in these equations can be found in
Appendix C.
\par
This  equation is not  very inspiring as it stands.
 But when we try to compute the
noise in first order to perturbation theory according to equation
(2.11) we can verify that
\begin{equation}
{\bf P} [\hat{A}, \hat{V}] = [\hat{A}, \hat{V}]
\end{equation}
which means that the lowest order term in the perturbation expansion
of the noise term is vanishing. Hence in a weak coupling regime one can
consider only the first order to $\lambda$ which gives an autonomous
and time-independent set of equations, for our relevant variables. The
one-point functions evolve freely and drive the corrsponding values of
the two-point correlation dunctions.
\begin{eqnarray}
 \frac{\partial}{\partial t} a({\bf k}) + i \omega_{\bf k} a({\bf k}) = 0  \hspace{4cm}
\\
\frac{\partial}{\partial t} Z({\bf k}) =  \frac{\lambda}{4} \left(
\frac{2}{\omega_{\bf k}} +  \frac{1}{(\omega_{\bf k} \omega_{{\bf k}/2})^{1/2}} \right) [a({\bf k})
+ a^*({\bf k})] \\
\frac{\partial}{\partial t} G({\bf k}) = - 2 i \omega_{\bf k} G({\bf k})  + \frac{
\lambda}{4} \left( 
\frac{2}{\omega_{\bf k}} +  \frac{1}{(\omega_{\bf k} \omega_{{\bf k}/2})^{1/2}} \right) [a({\bf k})
+ a^*({\bf k})]
\end{eqnarray}

\section{Conclusions and remarks}
\par
The techniques we have employed in this paper have given as a picture
for the evolution of relevant variables, when the coarse-graining
operation consists in the truncation of the Schwinger-Dyson hierarchy
of n-point 
functions. 
\par
One of the great difficulties in such considerations is the
complicated expressions we get for our equations in the end. It seems
that it is very difficult to find a regime in a field theory where the
dynamics would be Markovian. This essentially means that noise should
be with good aproximation ``white'' and in the autonomous part of the
dynamics one should have no time-dependent coefficents. It seems
unlikely that we can obtain Markovian evolution for a generic state of
the system. In any case we should expect it when the field is in a
state of partial (local) equilibrium \cite{Zeh}. This regime can still
be studied using our techniques, but it might be that a different
choice of coarse-graining projector might be more of use. The
Kawasaki-Gunton and the Mori projector \cite{RaMi} might prove more
convenient when dealing with this regime.
\par
Another avenue to explore towards obtaining Markovian equations is to
consider non-unitary dynamics for the evolution of the total
system. This might come from a contact with a heat bath or through the
interaction with other ignored degrees of freedom (a supermassive
field or gravitons for the case of cosmology). 
\par
As far as the noise  is concerned, we should stress that the
Zwanzig method allows to derive the noise term in the evolution
equations solely from the knowledge of the initial state of the
system. The comparison of its strength with the size of the terms
entering the evolution equations offers a good criterion ( though
rather heuristic) for the classicalization of the variables under
study. Remember, that noise should be strong enough to decohere but
weak enough to allow for predictability and not covering up the
effects of the potential. Only a particular class of initial states
offers this possibility.
\par
 Finally, we should make some remarks concerning the classical domain
in generic field theories. The techniques developed in this paper do
provide a useful tool for dealing with the emergence of classical
behaviour. Still, it is my belief, that concrete understanding of the
quantum to classical transition requires in addition, employment of the
 conceptual technical tools of the decoherent histories approach to quantum
mechanics. To obtain a complete and rigorous 
characterization of the classical domain ( like for instance
\cite{Ana,Omn,DoHa}) one needs to construct 
the decoherence
functional for coarse grained correlation histories in a manageable
computationally form. This is currently under investigation.

\section{Aknowledgements}
I would like to thank B. L. Hu for a stimulating discussion on
relevant issues.

\pagebreak
\begin{appendix}
\section{The index notation}
In the paper we have heavily used an index notation connected with the
normal-ordered form of an operator, which we describe in detail here.
\par
For reasons of symmetry in our expressions we prefer to work using the
creation and annihilation operators in the configuration space instead
of the momentum as is usual. Hence we write $\hat{a}(x)$ and
$\hat{a}^{\dagger}({\bf x})$ . They are related to the standard operators in
momentum space by
\begin{equation}
\hat{a}({\bf x}) = \int \frac{dk}{(2 \omega_{\bf k})^{1/2})} e^{-i{\bf k}{\bf x}} \hat{a}({\bf k})
\end{equation}  
We denote $\hat{a}({\bf x})$ by $\hat{a}^a$ (index up) and
$\hat{a}^{\dagger}( {\bf x})$ by $\hat{a}^{\dagger}_a$ (index down).
 To any function or distribution assign an abstract index
 to each of its arguments. The index is lower or upper  according to
whether the corresponding argument is
integrated out with an $\hat{a}$ or an $\hat{a}^{\dagger}$
respectively. Hence the operator 
\begin{equation}
\int dx_1 dx_2 dx_3 K( {\bf x}_1, {\bf x}_2; {\bf x}_3) \hat{a}^{\dagger}( {\bf x}_1)
\hat{a}^{\dagger}( {\bf x}_2) \hat{a}( {\bf x}_3)
\end{equation}
will be represented as  
\begin{equation}
\hat{a}^{\dagger}_a \hat{a}^{\dagger}_b K^{ab}_{\;\;\;\;c} \hat{a}_c
\end{equation}
We can easily verify that lowering a single index corresponds to
changing the argument in the distribution from${\bf x}$ to $- {\bf x}$, and
inversion of all indices amounts to complex conjugation.

\section{Useful formulae}
Here we list here a number of expressions of which we make use in the paper.
\par
The free Hamiltonian can be writen
\begin{equation}
\hat{H_0} = \frac{1}{2} \int dx dx' \hat{a}^{\dagger} ({\bf x}) h({\bf x},{\bf x}') \hat{a}({\bf x}')
\end{equation}
 with 
\begin{equation}
h({\bf x},{\bf x}') = \int dk e^{-i{\bf k}({\bf x}-{\bf x}')} \omega_{\bf k}
\end{equation}
The   evolution operator $\hat{U}_0(t) =
e^{-it\hat{H_0}}$ reads
\begin{equation}
\hat{U_0} (t) = :\exp \left[ \int dx \hat{a}^{\dagger}({\bf x})
(\Delta({\bf x}-{\bf x}';t) - \delta({\bf x}-{\bf x}'))
\hat{a}({\bf x}') \right]:
\end{equation}
where
\begin{equation}
\Delta({\bf x}-{\bf x}';t) = \int dk e^{-i  {\bf k}({\bf x}-{\bf x}')} e^{-i \omega_{\bf k} t}
\end{equation}
A coherent state is characterized by the square integrable function
$\alpha({\bf x})$ and is an eigenstate of the annihilation operator
$\hat{a}({\bf x})$. Under evolution of the free Hamiltonian we have
\begin{equation}
\hat{U}_0(t) |\alpha({\bf x}) \rangle = | \alpha({\bf x},t) \rangle
\end{equation}
where 
\begin{equation}
\alpha({\bf x},t) = \int dx' \Delta({\bf x}-{\bf x}';t) \alpha({\bf x}')
\end{equation}
\\
The operator 
\begin{equation}
\hat{V} = :\int dx \frac{g}{3!} \hat{\phi^3}:
\end{equation}
reads in the index notation
\begin{equation}
V(\alpha^*,\alpha) = \frac{g}{3!} \left( V_{abc} \hat{a}^a \hat{a}^b
\hat{a}^c + 3 \hat{a}^{\dagger}_a V^a_{\; \; bc} \hat{a}^b \hat{a}^c + 3 \hat{a}^{\dagger}_a
\hat{a}^{\dagger}_b V^{ab}_{\quad c} \hat{a}^c + \hat{a}^{\dagger}_a \hat{a}^{\dagger}_b \hat{a}^{\dagger}_c
V^{abc} \right)
\end{equation} 
with the correspondence
\begin{eqnarray}
V_{abc} & \leadsto \int \prod_{i=1}^{3} \frac{dk_i}{(2 \omega_{{\bf k}_i})^{1//2}}
e^{-i({\bf k}_1{\bf x}_1+{\bf k}_2 {\bf x}_2 +{\bf k}_3 {\bf x}_3)} (2 \pi)^3 \delta({\bf k}_1 + {\bf k}_2 + {\bf k}_3)
 \\
V^a_{ \; \; bc} & \leadsto \int \prod_{i=1}^{3} \frac{dk_i}{(2
\omega_{{\bf k}_i}^{1/2} )}
e^{-i(-{\bf k}_1{\bf x}_1+{\bf k}_2 {\bf x}_2 +{\bf k}_3 {\bf x}_3)} (2 \pi)^3 \delta({\bf k}_1 + {\bf k}_2 + {\bf k}_3)
 \\
V^{ab}_{\quad  c} & \leadsto \int \prod_{i=1}^{3} \frac{dk_i}{(2
\omega_{{\bf k}_i})^{1/2} }
e^{-i(-{\bf k}_1{\bf x}_1-{\bf k}_2 {\bf x}_2 +{\bf k}_3 {\bf x}_3)} (2 \pi)^3 \delta({\bf k}_1 + {\bf k}_2 + {\bf k}_3)
 \\
V_{abc} & \leadsto \int \prod_{i=1}^{3} \frac{dk_i}{(2 \omega_{{\bf k}_i})^{1/2}}
e^{i({\bf k}_1{\bf x}_1+{\bf k}_2 {\bf x}_2 +{\bf k}_3 {\bf x}_3)} (2 \pi)^3 \delta({\bf k}_1 + {\bf k}_2 + {\bf k}_3)
\end{eqnarray}
while the operator 
\begin{equation}
\hat{V} = :\int dx \frac{\lambda}{4!} \hat{\phi^4}:
\end{equation}
reads
\begin{eqnarray}
\hat{V} = \frac{\lambda}{4!} \left( V_{abcd} \hat{a}^a \hat{a}^b
\hat{a}^c \hat{a}^d + 4 \hat{a}^{\dagger}_a V^a_{\;\;bcd}\hat{a}^b
\hat{a}^c \hat{a}^d \right. \nonumber \\ 
\left. + 6 \hat{a}^{\dagger}_a \hat{a}^{\dagger}_b 
V^{ab}_{\quad cd} \hat{a}^c \hat{a}^d + 4  \hat{a}^{\dagger}_a
\hat{a}^{\dagger}_b  \hat{a}^{\dagger}_c V^{abc}_{\quad \; d}
\hat{a}^d + \hat{a}^{\dagger}_a \hat{a}^{\dagger}_b
\hat{a}^{\dagger}_c \hat{a}^{\dagger}_d \right)
\end{eqnarray}
Vith
\begin{eqnarray}
V_{abcd} \leadsto  \int \prod_{i=1}^{4} \frac{dk_i}{(2
\omega_{{\bf k}_i})^{1/2} } e^{-i({\bf k}_1{\bf x}_1+{\bf k}_2 {\bf x}_2 +{\bf k}_3 {\bf x}_3+{\bf k}_4{\bf x}_4)} \; (2
\pi)^3 \delta({\bf k}_1+{\bf k}_2 +{\bf k}_3 +{\bf k}_4)\hspace{1cm}
\\
V^a_{\;\;bcd} \leadsto  \int \prod_{i=1}^{4} \frac{dk_i}{(2
\omega_{{\bf k}_i})^{1/2} } e^{-i(-{\bf k}_1{\bf x}_1+{\bf k}_2 {\bf x}_2 +{\bf k}_3 {\bf x}_3+{\bf k}_4{\bf x}_4)} \; (2
\pi)^3 \delta({\bf k}_1+{\bf k}_2 +{\bf k}_3 +{\bf k}_4)
\hspace{1cm} 
\\ 
V^{ab}_{\quad cd} \leadsto  \int \prod_{i=1}^{4} \frac{dk_i}{(2
\omega_{{\bf k}_i})^{1/2} } e^{-i(-{\bf k}_1{\bf x}_1-{\bf k}_2 {\bf x}_2 +{\bf k}_3 {\bf x}_3+{\bf k}_4{\bf x}_4)} \; (2
\pi)^3 \delta({\bf k}_1+{\bf k}_2 +{\bf k}_3 +{\bf k}_4)
\hspace{1cm} 
\\
V^{abc}_{\quad \;\;d} \leadsto  \int \prod_{i=1}^{4} \frac{dk_i}{(2
\omega_{{\bf k}_i})^{1/2} } e^{-i(-{\bf k}_1{\bf x}_1-{\bf k}_2 {\bf x}_2 -{\bf k}_3 {\bf x}_3+{\bf k}_4{\bf x}_4)} \; (2
\pi)^3 \delta({\bf k}_1+{\bf k}_2 +{\bf k}_3 +{\bf k}_4)
\hspace{1cm} 
\\
V^{abcd} \leadsto  \int \prod_{i=1}^{4} \frac{dk_i}{(2
\omega_{{\bf k}_i})^{1/2} } e^{i({\bf k}_1{\bf x}_1+{\bf k}_2 {\bf x}_2 +{\bf k}_3 {\bf x}_3+{\bf k}_4{\bf x}_4)} \; (2
\pi)^3 \delta({\bf k}_1+{\bf k}_2 +{\bf k}_3 +{\bf k}_4)
\hspace{1cm} 
\end{eqnarray}
\section{The coefficients in equations (4.8-4.10)}
Here we give the expressions for the coefficients in equation (4.8-4.10)

\begin{eqnarray}
r({\bf k},{\bf k}';t) = \frac{3}{2}\frac{1}{\omega_{{\bf k}+{\bf k}'} \omega_{{\bf k}'}^2} \int_0^t d
\tau e^{-i \omega_{{\bf k}'}\tau} \sin(\omega_{\bf k} + \omega_{{\bf k}+{\bf k}'})\tau 
\nonumber \\
- \frac{1}{4} \frac{1}{\omega_{{\bf k}'}^2 \omega_{{\bf k}/2}} \frac{\cos
\omega_{\bf k}t - 1}{\omega_{\bf k}}
\\
s({\bf k},{\bf k}';t) = \frac{3}{2}\frac{1}{\omega_{{\bf k}+{\bf k}'} \omega_{{\bf k}'}^2} \int_0^t d
\tau \cos \omega_{{\bf k}'} \tau \sin (\omega_{\bf k} + \omega_{{\bf k}+{\bf k}'}) \tau
\nonumber \\
-\frac{1}{2} 
\frac{1}{\omega_{{\bf k}'}^2 \omega_{{\bf k}/2}} \frac{\cos
\omega_{\bf k}t - 1}{\omega_{\bf k}}
\\ \\
D_1({\bf k};t) = 3 u({\bf k};t) - 2 u'({\bf k};t)  \\
D_2({\bf k};t) = - 3 u^*({\bf k};t) - 2 u'({\bf k};t) \\
D_3({\bf k};t) = 3 [u({\bf k};t)+u^*({\bf k};t)] - 4 u'({\bf k},t) \\
\\
   u({\bf k};t) = \frac{1}{4} \frac{1}{\omega_{\bf k}} \int 
\frac{dk_1}{\omega_{{\bf k}_1} \omega_{{\bf k}+{\bf k}_2}} \int_0^t d \tau e^{-i \omega_{\bf k}
\tau} \sin(\omega_{{\bf k}_1} + \omega_{{\bf k}+{\bf k}_1}) \tau 
\\
u'({\bf k};t) = \frac{1}{2} \frac{1}{\omega_{\bf k}^3 \omega_{{\bf k}/2}} \left( \cos
\omega_{\bf k} t - 1 \right) \\
\\
 K_1({bf {\bf k}},{\bf k}';t) = \frac{3}{4} \frac{1}{\omega_{{\bf k}+{\bf k}'} \omega_{{\bf k}'}^2} \frac{
\cos(\omega_{\bf k} + \omega_{{\bf k}+{\bf k}'} - \omega_{{\bf k}'}) t - 1}{\omega_{\bf k} +
\omega_{{\bf k}+{\bf k}'} - \omega_{{\bf k}'} } \nonumber \\
-\frac{1}{2} 
\frac{1}{\omega_{{\bf k}'}^2 \omega_{{\bf k}/2}} \frac{e^{-i
\omega_{\bf k}t} - 1}{\omega_{\bf k}} \\
K_2({\bf k},{\bf k}';t) = \frac{3}{2} \frac{1}{\omega_{{\bf k}+{\bf k}'} \omega_{{\bf k}'}^2} 
\int_0^t e^{-i (\omega_{\bf k} + \omega_{{\bf k}'})\tau \cos(\omega_{{\bf k}+{\bf k}'}\tau}
\nonumber \\
-\frac{1}{2} 
\frac{1}{\omega_{{\bf k}'}^2 \omega_{{\bf k}/2}} \frac{e^{-i
\omega_{\bf k}t} - 1}{\omega_{\bf k}} \\
K_3({\bf k},{\bf k}';t) =  \frac{1}{2}\frac{1}{\omega_{{\bf k}+{\bf k}'} \omega_{{\bf k}'}^2} 
\frac{
\cos(\omega_{\bf k} + \omega_{{\bf k}+{\bf k}'} + \omega_{{\bf k}'}) t - 1}{\omega_{\bf k} +
\omega_{{\bf k}+{\bf k}'} + \omega_{{\bf k}'} } \\ 
\\ 
L_1({\bf k};t) = -\frac{3i}{4}  \frac{1}{\omega_{\bf k}} \int 
\frac{d{\bf k}_1}{\omega_{{\bf k}_1} \omega_{{\bf k}+{\bf k}_2}} \int_0^t d \tau e^{-i \omega_{\bf k}
\tau} \cos(\omega_{{\bf k}_1} + \omega_{{\bf k}+{\bf k}_1}) \tau \nonumber \\
- \frac{1}{\omega_{{\bf k}'}^2 \omega_{{\bf k}/2}\omega_{\bf k}} \left( \cos
\omega_{\bf k} t - 1 \right)\\
L_2({\bf k};t) = - \frac{1}{\omega_{\bf k}^3 \omega_{{\bf k}/2}} \left( \cos
\omega_{\bf k} t - 1 \right) \nonumber \\
+ \frac{3}{4} \frac{1}{\omega_{\bf k}} \int \frac{d{\bf k}_1}{\omega_{{\bf k}_1}
\omega_{{\bf k}+{\bf k}_1}} \int_0^t d \tau e^{i \omega_{{\bf k}}\tau} \sin \omega_{{\bf k}_2
+ {\bf k}} \tau  
\end{eqnarray}
\end{appendix}

\begin{thebibliography}{}
\bibitem{CaHu}  B. L. Hu and E. Calzetta,  ``Decoherence of
correlation histories'', 
in {\sl Directions in General Relativity, vol II:
Brill Festschrift},edited by B.L Hu and T. A. Jacobson (Cambridge
University Press, Cambridge University Press, Cambridge 1994).
 9302013
9501040; `` Correlations, Decoherence, Dissipation and Noise in
Quantum Field Theory'', in {\sl Heat Kernel Techniques and Quantum
Gravity}, edited by S. A. Fulling (Texas AM Press, College Station 1995)
 gr-qc 9501040.

\bibitem{Ber}  F.\ A.\ Berezin, {\sl The Method of Second
Quantization}, (Academic Press, New York, 1966).

\bibitem{Zeh} H.\ D.\ Zeh, { \sl The Physical Basis of the Direction of
Time} (Springer Verlag, Berlin, 1989).

\bibitem{Ana} C.\ Anastopoulos, {\sl Phys. Rev.}{ \bf E53}, 4711 (1996).

\bibitem{Omn} R.\ Omn\`es, {\sl J. Stat. Phys. }{\bf 57}, 357 (1989);
{\sl The Interpretation of Quantum 
Mechanics} 
( Princeton University Press, Princeton,
1994) 

\bibitem{DoHa} H.\ F.\ Dowker and J.\ J.\ Halliwell, {\sl Phys. Rev. D}{\bf 46}, 1580 (1992).
\bibitem {FeVe} R.\ P.\  Feynman and A.\ R.\ Hibbs, Quantum Mechanics and
path integrals (McGraw-Hill, New York, 1965) ; R.\ P.\ Feynman and F.\
 L.\ Vernon, {\sl Annals of Physics} {\bf 24}, 118 (1963).

\bibitem {CaLe} A.\ O.\ Caldeira and A.\ J.\ Leggett, {\sl Physica
A}{\bf 121}, 587 (1983).

\bibitem{HuMa} B.\ L.\ Hu and A.\ Matacz, {\sl Phys. Rev. D}{\bf49},
6612 (1994).

\bibitem{LiWe} K.\ Lindenberg and B.\ J.\ West, {\sl The
non-equilibrium statistical mechanics of open and closed systems},(VHC
Publishers, New York 1990) 

\bibitem{AnZu} J. R. Anglin and W. H> Zurek, {\sl Decoherence of
quantum fields: Pointer states and Predictability}, gr-qc 9510021.

\bibitem{CaHu2} B. L. Hu and E. Calzetta, {\sl Phys. Rev.} {\bf D52},
6770(1995).

\bibitem{CaHu3} B. L. Hu and E. Calzetta, {\sl Phys. Rev.}{\bf D35},
495 (1987); {\sl Phys. Rev} {\bf D} 40, 656 (1989).

\bibitem{RaMi} J.\ Rau and B.\ Muller, {\sl From reversible quantum
microdynamics to irreversible quantum transport}, nucl-th 9505009.


\bibitem {GeHa} M.\ Gell-Mann and J.\ B.\ Hartle, in {\sl Complexity, Entropy 
and the Physics of Information}, edited by W.\ Zurek, ({\sl Addison
Wesley,Reading} (1990));  
{\sl Phys. Rev. D}{\bf47}, 3345 (1993).
\end{thebibliography}
\end{document}